\documentstyle[12pt]{article}
\newcommand{\iif}{\infty}
\newcommand{\ts}{\times}
\newcommand{\newref}[1]{\par \noindent $\!\!\!\!\!^{#1}$}
\newcommand{\newreff}[1]{\par \noindent $\!\!\!\!\!\!\!\!^{#1}$} 
\newcommand{\be}{\begin{equation}}
\newcommand{\ee}{\end{equation}}
\title{WKB --- Not So Bad After All}
\author{Chung-Sang Ng \\ {\em Physics Department, Auburn University, Auburn, AL 36849}}
\date{July 6, 1992}   
\begin{document}
\maketitle
\baselineskip 25pt
\vspace{1in}
\noindent
It was found recently that tunneling probabilities
over a barrier is
roughly twice as large as that given by standard WKB formula.  Here
we explained how this come from and showed that WKB method does
give a good approximation over almost
entire energy range provided that we use appropriate connection
relations. 

\vspace{1in}
\vfill

\section*{I.  Introduction}

WKB method was first invented by Jeffreys$^{1}$ and was applied to solve
Schr\"{o}dinger equation by Wentzel, Kramers and Brillouin$^{2}$.  It is a
powerful tool and has many applications, e.~g.~, waves in a
inhomogeneous plasma$^{3}$.  Consider a
second order ordinary differential equation:
\be
\psi''(x)+k^{2}(x)\psi(x) = 0,
\ee
where prime denotes differentiation with respect to argument.  
The function
$k^{2}$ may be negative for some $x$. 
 In this case, we define $\kappa^{2}(x) \equiv - k^{2}(x)$. 
WKB method approximates solutions of (1) by combinations of
\be
 \left.\exp\left(\pm i \int^{x}{k(y)\,dy}\right)\right/\sqrt{k(x)}\ ,
\ee
provided that
\be
 \left|k'/k^{2}\right| \ll 1
\ee
is satisfied.  If (3)
is not satisfied over a range (or at a point), then we need to connect
WKB solutions of both sides appropriately by some connection relations. 
One common situation where (3) breaks down is that $k^{2} \approx 0 $
near a turning point. 

In a recent paper$^{4}$, 
a quantum mechanical tunneling example is considered.  The
Schr\"{o}dinger equation in this one dimensional problem 
can be represented by (1) with
\be
  k^{2}(x) = \left\{ \begin{array}{ll}
                   E - \alpha x^{2} & \mbox{for $ 0 \leq x \leq b $} \\
                   E                & \mbox{otherwise,}
                  \end{array}
           \right. 
\ee
where $b = \sqrt{E_{b}/\alpha}$ and $E$, $E_{b}$ and $\alpha$ are constants. 
$E$ may be assigned the physical meaning of energy. 
It was shown that, for $E < E_b$,  
the true (numerical) tunneling probabilities $T_{n}$ (see
Appendix for numerical method) are roughly twice as large as the well known
WKB tunneling factor:
\be
T_{w} = e^{-2\int_{a}^{b}{\kappa(x)\, dx}}=e^{\frac{E}{\sqrt{\alpha}}
\left\{\ln[\beta+\sqrt{\beta^{2}-1}]-\beta\sqrt{\beta^{2}-1}\right\}}, 
\ee
where $a = \sqrt{E/\alpha}$ and $\beta = \sqrt{E_{b}/E}$.  Values of $T_{w}$
and $T_{n}$ for different $E$ are shown in Table~I and Table~II. 

This result was unexpected to us at first.  
After a little survey over some standard textbooks, 
it was found that two books did warn
about such an error$^{5,6}$, although without elaborations.  Another 
book did an example
with correct connection relation$^{7}$,  but did not give general discussions. 
Two other books used 
the standard tunneling factor without warning$^{8,9}$, 
although one of them indicated 
that it is just a rough approximation$^{9}$.  In this case,  
we have reasons to
assume that the restrictions in applying the standard WKB formula are not
very well known to students.  Therefore, we would like to
show how to apply WKB method to solve a tunneling problem more carefully. 

In section II,  we briefly review the derivation of the standard WKB tunneling
factor using standard connection relations, which are also useful for our
discussions later,  over a typical turning point.  In section III,  the tunneling
factor is corrected when one turning point is a sharp edge.  In
section IV,  further correction is made when the tunneling probability is not
small.  Section V considers the case that $E \rightarrow 0$.  Section VI deals with the case that $ E \approx E_{b}$.  
Finally section VII treats the case that
$E > E_{b}$.  It will be shown that WKB method does give good approximation
to the tunneling probability if we use different connection relations
for different energy range appropriately. 

\section*{II.  Standard WKB tunneling formula}

Consider (1) with $x$ near a turning point $x = 0$.  Assume that 
$k^{2}(x)$ can
be approximated by:
\be
   k^{2}(x) \approx \left.\frac{dk^{2}}{dx}\right|_{x=0}x\, . 
\ee
Let $\left(dk^{2}/dx\right)_{0} > 0 $ for the time being.  
Solutions of (1) with $k^{2}$ given by (6) 
can be written as combinations of Airy's functions$^{10}$ \,  
$\mbox{Ai$(-\lambda x)$}$ and $\mbox{Bi$(-\lambda x)$}$ with $\lambda \equiv
\left|\left(dk^{2}/dx\right)_{0}\right|^{1/3}.$
Then the asymptotic behaviors of Ai and Bi give the standard
WKB connection relations$^{10}$:
\be
\begin{array}{ccccc}
-\infty & \longleftarrow & x & \longrightarrow & \infty \vspace{2.5mm}\\
\frac{1}{2\sqrt{\kappa}}e^{-\int_{x}^{0}{\kappa\,  dy}} &  \longleftarrow &
\sqrt{\frac{\pi}{\lambda}}\mbox{Ai$(-\lambda x)$}&\longrightarrow &
\frac{1}{\sqrt{k}}\sin(\int_{0}^{x}{k\,  dy}+\frac{\pi}{4}) \vspace{2.5mm}\\
\frac{1}{\sqrt{\kappa}}e^{\int_{x}^{0}{\kappa\,  dy}} &  \longleftarrow &
\sqrt{\frac{\pi}{\lambda}}\mbox{Bi$(-\lambda x)$}&\longrightarrow &
\frac{1}{\sqrt{k}}\cos(\int_{0}^{x}{k\,  dy}+\frac{\pi}{4}). \vspace{2.5mm} \\
\end{array}
\ee
For $\left(dk^{2}/dx\right)_{0} < 0 $, we only need to change $\lambda$
to $-\lambda$ in (7),  interchange the limits of integrations and interchange
left and right hand sides.  These standard connection relations are valid
only if the turning point is smooth enough, i.e., we require (6) to be valid 
up to some $x$
where (3) is also satisfied.  This is not always possible$^{3}$. 

Consider a barrier with $k^{2} > 0 $ for $x \leq a$ and $x \geq b$, 
$k^{2} < 0 $ for $ a \leq x \leq b $ and that WKB condition (3) is valid
over the whole range of $x$ except near the two smooth turning points
$a$ and $b$. Then for a incident wave from $x = -\infty$ direction,  we have
only outgoing wave for $x > b $:
\be
\psi = \frac{1}{\sqrt{k}}e^{i\int_{b}^{x}{k\,  dy}}. 
\ee
By (7),  $\psi$ connects to:
\be
\psi=\left[\frac{1}{\sqrt{\kappa}}e^{\int_{x}^{b}{\kappa\,  dy}}
      + \frac{i}{2\sqrt{\kappa}}e^{-\int_{x}^{b}{\kappa\,  dy}}\right]
      e^{-i\pi/4}
\ee
for $a < x < b$. Then for $x$ close to $a$, 
\be
\psi \approx \frac{e^{I}}{\sqrt{\kappa}}e^{-\int_{a}^{x}{\kappa\,  dy}-i\pi/4}, 
\ee
where $I \equiv \int_{a}^{b}{\kappa\,  dy}$ and is assumed large.  By (7) again, 
$\psi$ connects to
\be
\psi = 2\frac{e^{I-i\pi/4}}{\sqrt{k}}\sin\left(\int_{x}^{a}{k\,  dy}+\pi/4\right)
\ee
for $x \leq a$. Since the sine term is just a combination of incoming and
reflected wave with equal intensity,  we have reflection coefficient
$R_{w} \approx 1$ and tunneling probability $T_{w} = e^{-2I}$.  This is
how (5) come from. 

From the above derivation,  we see that in order to apply the standard
WKB tunneling factor (5), three conditions must be satisfied :
\begin{description}
\item[(i)\, \, ] there are two and only two turning points;
\item[(ii)\, ] (3) is valid except near the two turning points where $k^{2}$
            can be approximated by (6); and
\item[(iii)] $T_{w} \ll 1$ . 
\end{description}
We immediately see that the potential given by (4) does not satisfy (ii)
since $k^{2}$ is discontinuous at $x = b$. This is the main reason for the
factor of two error.  Also,  for larger $E$,  $T_{w}$ is not very small,  (iii)
is also violated. 

\section*{III.  Sharp turning point correction}

Let us now consider $k^{2}$ with a discontinuity at $x = b$, 
so that $k^2= -\kappa^2_{b} < 0$ at $b- 0^{+}$ and $k^2 = k^2_{b} > 0$ at
$b + 0^{+}$. We still assume that the 
WKB condition (3) remains valid except at $b$ and
near the smooth turning point $x = a$.  For $x > b$,  there is only
outgoing wave (8).  It connects to
\be
\psi=\frac{1}{2\sqrt{\kappa}}\left\{\left[\sqrt{\frac{\kappa_{b}}{k_{b}}}
-i\sqrt{\frac{k_{b}}{\kappa_{b}}}\, \right]e^{\int_{x}^{b}{\kappa\,  dy}}
      + \left[ \sqrt{\frac{\kappa_{b}}{k_{b}}}
+i\sqrt{\frac{k_{b}}{\kappa_{b}}}\, \right]e^{-\int_{x}^{b}{\kappa\,  dy}}\right\}
\ee
for $ a < x \leq b$. 
We can check whether (12) is correct by substituting $x = b$ into (8) and (12). 
We should see that $\psi$ and $\psi'$ are indeed continuous. 
Similar to steps (9) to (11) and still assuming $I$ to be large, 
the corrected tunneling probability can be found as:
\be
   T_{1} = \frac{4\kappa_{b}k_{b}}{\kappa_{b}^{2}+k_{b}^{2}}T_{w}. 
\ee
For $k^{2}$ given by (4), 
\be
    T_{1} = \frac{4\sqrt{E_{b}/E - 1}}{E_{b}/E}T_{w}. 
\ee
It can be easily shown by (13) that $T_{1} \leq 2T_{w}$.  The equal sign
holds for $k_{b} = \kappa_{b}$.  This explains the factor of two 
found by ref.\ 4,  Some values of $T_{w}$ and $T_{1}$
are shown in Table~I to compare with true (numerical) tunneling
probabilities $T_{n}$.   Parameters $\alpha = 0.040965,  
E_{b}=1.2776$ were chosen so that we may compare with Table~I of ref.\ 4.  
We see that $T_{1}$ is much closer to $T_{n}$ from $E \approx 0.2 $ to 
$E \approx 0.6$ while $T_{w}$
is nearly a factor of two smaller. 
However,  for large $E$ when $T_{w}$ lager than
0.2,  $T_{1}$ also fails since we assumed $T_{w}$ to be small in the above
derivation. This error will be corrected in next section.


\section*{IV. Finite $T_{w}$ correction}

First, let us consider the correction of $T_{w}$ itself. If we keep both
terms in (9),  then $\psi$ will connects to
\be
\psi=\frac{1}{\sqrt{kT_{w}}}\left[2\sin(\theta)+\frac{i}{2}T_{w}\cos(\theta)
       \right]e^{-i\pi/4}
\ee
for $x < a$ where $\theta \equiv \int_{x}^{a}{k\, dy}+\pi/4$. Then by finding
the coefficient of incoming term,  the corrected tunneling probability
can be found as:
\be
T_{w2} = T_{w}/\left(1+T_{w}/4\right)^{2}.
\ee
Similarly,  keeping both terms in (12) gives correction of $T_{1}$ :
\be
T_{2} = T_{1}/\left(1+T_{1}/2+T_{w}^{2}/4 \right).
\ee
Some values of $T_{2}$ are also shown in Table~I and Table~II. We see that
$T_{2} \rightarrow T_{1}$ as $T_{w} \rightarrow 0$. But for large $T_{w}$, 
$T_{2}$ differs from $T_{1}$ quite a lot and gives better approximation to
$T_{n}$ up to $E \approx 1$,  or $T_{w} \approx 0.65$,  which is quite large.
However, for $E \rightarrow E_{b}$,  $T_{2}$ also fails. For example, 
for $E = E_{b}$,  both $T_{1}$ and $T_{2}$ equal to zero while $T_{n}$ is
actually quite large. The reason for this error is that we assumed that WKB
condition (3) is satisfied up to $x = b$. This is not true 
for $E \rightarrow E_{b}$,  since $\kappa$ is small,  even for $x = b$, 
while $\kappa'$ is not small. So we
cannot use (13) or (17) for $E$ close to $E_{b}$. Another way to apply
WKB method for this case is discussed in section VI.

In the other extreme,  the fact that for $E \rightarrow 0$, $T_{w}
\rightarrow $finite value while $T_{n} \rightarrow 0$, is another limitation
of the standard WKB tunneling factor$^{4}$. Although $T_{1}$ and $T_{2}$
give the right value,  i.e.\ zero,  at $E = 0 $ as shown in Table I,  the
dependence of $E$ is wrong as we can see from Table II. We see that
$T_{n}\propto E $ as $E \rightarrow 0$ while $T_{2} \approx T_{1}
\propto \sqrt{E}$. So,  there are extremely large error between $T_{n}$
with $T_{1}$ or $T_{2}$. We will discuss this case in the next section.


\section*{V. For $E \rightarrow 0$}

Consider the general case first. Let $k^{2} = E \rightarrow 0$
for $ x \leq a $. By (12) and connection formula (7),  we know
\be
\psi \approx \frac{1}{\sqrt{T_{w}}}\left[\sqrt{\frac{\kappa_{b}}{k_{b}}}
-i\sqrt{\frac{k_{b}}{\kappa_{b}}}\, \right]\sqrt{\frac{\pi}{\lambda}}
\mbox{Ai$(\lambda x)$}
\ee
for $x \rightarrow a^{+}$, where $\lambda^{3} 
\equiv -\left(dk^{2}/dx\right)_{x = a}$. 
Again,  $T_{w}$ small was assumed, although this assumption can be
removed if higher accuracy is desired. Now assume
\be
   \psi = Ce^{i\sqrt{E}x} + De^{-i\sqrt{E}x}
\ee
for $x < a \rightarrow 0$, where $C$ and $D$ are constants.
Matching (18) and (19) at $x = 0$ and
using the fact that $E \rightarrow 0$,  we got the tunneling probability:
\be
T = \frac{1}{|C|^{2}\sqrt{E}} = \frac{\sqrt{E}}{\pi\lambda
\left[\mbox{Ai}'(0)\right]^{2}} T_{1}.
\ee
Note that power series expansions for Airy's functions are$^{10}$:
\be
\begin{array}{ccccc}
\mbox{Ai$(z)$} & = &c_{1}f(z) & - &c_{2}g(z), \vspace{2.5mm} \\
\mbox{Bi$(z)$} & = &\sqrt{3}[c_{1}f(z)& + &c_{2}g(z)], 
\end{array}
\ee
where
\begin{eqnarray*}
f(z) & = &1+\frac{z^{3}}{2\cdot3}\left\{1+\frac{z^{3}}{5\cdot6}\left[
1+\frac{z^{3}}{8\cdot9}\left(1+\cdots\right)\right]\right\}, \vspace{1.5mm} \\
g(z) & = &z \cdot\left(1+\frac{z^{3}}{3\cdot4}\left\{
1+\frac{z^{3}}{6\cdot7}\left[1+\frac{z^{3}}{9\cdot10}
\left(1+\cdots\right)\right]\right\}\right), \vspace{1.5mm}\\
c_{1} & = &0.355028, \\
c_{2} & = &0.258819.
\end{eqnarray*}
So, Ai$'(0) = - c_{2}$. In the derivation of (20),  we assumed $\lambda$
finite and $k^{2}$ smooth enough near $x = a$. However,  for $k^{2}$ given
by (4), $\lambda = (4\alpha E)^{1/6} \rightarrow 0$.  This means that
$T \propto E^{5/6}$. This is not correct since $T_{n} \propto E$. 
The reason is that as $E \rightarrow 0$,  $\left(dk^{2}/dx\right)_{a}
\rightarrow 0$ while $\left(d^{2}k^{2}/dx^{2}\right)_{a} = -2\alpha$ is
constant.  This means that (6) is not valid and we cannot approximate
solutions by Airy's functions.  In order to correct this,  we need to consider
specifically the potential given by (4).  Now,  as $E \rightarrow 0$, 
$k^{2} \rightarrow -\alpha x^{2}$ for $ 0 \leq x \leq b $.  Use a change
of variable $y = \alpha^{1/4} x$,  (1) becomes:
\be
\psi_{yy} - y^{2}\psi = 0. 
\ee
This equation can be solved by so called parabolic cylinder 
functions$^{10}$. \ Actually,  we may use these functions to solve (1)
with $k^{2}$ given by (4) exactly and write the tunneling probability
in closed form.  This is out of the scope of this paper.  So let us
assume that we do not know these functions.  We will see that we 
do not need to solve (22) exactly.  Instead,  we may study it by
WKB method! First,  note that $\left(dy/dy
\right)/y^{2} = 1/y^{2} \ll 1 $ as $y \rightarrow \infty$,  i.e.\  the
WKB condition (3) is satisfied asymptotically. So (22) has asymptotic
solutions given by (2) with $k^{2} = -y^{2}$. Let us define two solutions of (22)
by their asymptotic behaviors as $y \rightarrow \infty$ :
\be
\begin{array}{ccccc}
\mbox{Aj$(y)$} & \rightarrow &\frac{1}{\sqrt{y}}e^{-\int_{0}^{y}{z^{2}\, dz}}
& = & \frac{1}{\sqrt{y}}e^{-y^{2}/2},  \vspace{2.5mm} \\
\mbox{Bj$(y)$} & \rightarrow &\frac{1}{\sqrt{y}}e^{\int_{0}^{y}{z^{2}\, dz}}
& = & \frac{1}{\sqrt{y}}e^{y^{2}/2} .\\
\end{array}
\ee
Then we may write the solution for $0 \leq x \leq b$ as
\be
\psi = \left[\mbox{Bj}'(\eta)\mbox{Aj}(y)
- \mbox{Aj}'(\eta)\mbox{Bj}(y)\right]\left/2\right., 
\ee
where $\eta \equiv \alpha^{1/4} b$, \,  in order to match with an outgoing
solution $e^{i\sqrt{E}x} \rightarrow 1$ for $x \geq b$. Note that we have 
used the
fact that the Wronskian $W = \mbox{Aj}\mbox{Bj}'-\mbox{Aj}'\mbox{Bj}
= 2 $ and that $E \rightarrow 0$. Using similar steps as (19) to (20)
and neglecting \mbox{Aj$'(\eta)$} as compared with \mbox{Bj$'(\eta)$},  
the tunneling probability can be found as:
\be
T_{0} =16E\left/\left\{\alpha^{1/2}\left[\mbox{Aj}'(0)
\mbox{Bj}'(\eta)\right]^{2}\right\}\right.  . 
\ee
This gives the correct $E$ dependence since $\alpha$ and $b$ are
independent of $E$.  To find \mbox{Bj$'(\eta)$}, we may use (23) as
a first approximation:
\be
\mbox{Bj}'(y) \approx \sqrt{y}e^{y^{2}/2}. 
\ee
So, 
\be
T_{0} = \frac{16E}{\alpha^{1/4}E_{b}^{1/2}\left[\mbox{Aj}'(0)\right]^{2}}T_{w}. 
\ee
This gives the dependence on $E$, $\alpha$ and $E_{b}$ since \mbox{Aj$'(0)$}
is only a constant. 
In order to compare $T_{0}$ with $T_{n}$ numerically,  we need to know
\mbox{Aj$'(0)$}, which can be found by numerical integration of (22). 
I found$^{11}$ $\mbox{Aj$'(0)$} \approx -0.9777$. Using this,  some values of $T_{0}$
are shown in Table~II.  We can see that they are quite close, although there
are more than 10\% difference.  The main error can be shown to be due to
the approximation in (26). To see this,  we note that the factor before
the exponential function in (23) should actually be an asymptotic series$^{10}$. 
The series of Bj can be found by requiring cancellations between terms
when it is put into (22):
\be
\mbox{Bj}(y) \longrightarrow \frac{e^{y^{2}/2}}{\sqrt{y}}\left(1+\frac{3}{4\cdot4
y^{2}}\left\{1+\frac{5\cdot7}{8\cdot 4y^{2}}\left[
1+\frac{9\cdot11}{12\cdot 4y^{2}}\left(1+\cdots\right)\right]\right\}\right)
.
\ee
This series can be evaluated numerically up to a term with smallest
magnitude.  This brought a factor of 1.133 to $T_{0}$ in Table~II, 
e.g.\ $T_{0}$ changed from $5.97 \times 10^{-10}$ to $6.76 \times
10^{-10}$ for $E = 10^{-8}$. This is very close to $T_{n}$ which is
$6.78 \times 10^{-10}$.

The dependence of $\alpha$ and $E_{b}$ in (27) were also be verified
numerically by choosing different $\alpha$ and $E_{b}$. In general, 
as $\alpha$ decreases or $E_{b}$ increases, i.e.\ $T_{w} \rightarrow 0$, 
the difference between $T_{0}$ and $T_{n}$ decreases. The fact that we
can find out the dependence of the tunneling probability on $E$, $\alpha$
and $E_{b}$ by using WKB idea without solving (22) shows how powerful
WKB method may be if applied correctly.


\section*{VI. For $E \approx E_{b}$}

For $E \approx E_{b}$,  (14) and (17) no longer give values close to $T_{n}$ as we can
see from Table~I. However,  we still can calculate the tunneling probability
by WKB method. Now,  $b-a$ is small,  we may approximate (4) by
$k^{2} \approx -\lambda^{3}(x-a)$ for $x$ in a region near $a$, including
$b$,  where $\lambda ^{3} = 2(\alpha E)^{1/2}.$ Then for this region,  the
solution is given by:
\be
\psi = \pi e^{i\sqrt{E}b}\left\{\left[\mbox{Bi}'(\xi)-\frac{i}{\lambda}
\sqrt{E}\mbox{Bi}(\xi)\right]\mbox{Ai}\left(z\right)
+ \left[-\mbox{Ai}'(\xi)+\frac{i}{\lambda}
\sqrt{E}\mbox{Ai}(\xi)\right]\mbox{Bi}\left(z\right)\right\}, 
\ee
where $z \equiv \lambda(x-a)$, $\xi \equiv \lambda(b-a)$, in order to
match with an outgoing solution $e^{i\sqrt{E}x}$ for $x \geq b$.
Using standard connection formula (7),  we may connect it to WKB type
solutions (2) for $x \rightarrow -\infty$. Then,  by grouping the
coefficient of the incoming terms,  we found the tunneling probability:
\be
T_{b} = \frac{4\lambda\sqrt{E}/\pi}
{\left[\mbox{Bi}(\xi)\sqrt{E}-\lambda\mbox{Ai}'(\xi)\right]^{2}
+\left[\mbox{Ai}(\xi)\sqrt{E}+\lambda\mbox{Bi}'(\xi)\right]^{2}}\, . 
\ee
The Airy's functions can be evaluated by power series expansions (21). 
Some values of $T_{b}$ are shown in Table~I for $E \leq E_{b}$ and
in Table~III for $E > E_{b}$.  We see that $T_{b}$ gives a quite good
approximation to $T_{n}$ for $|E-E_{b}| < 0.7$. The range of
validity for this approximation is surprisingly large
at first sight.  However,  if we remember that the potential is
proportional to $x^2$ so that although $\left|E-E_{b}\right|$ is
not small, $|a-b|$ may be small enough for the approximation to work. 

\section*{VII.  For $E > E_{b}$}

For the case that $E$ larger than $E_{b}$, $T_{w} = 1$ while the true
``tunneling probability''(it may be better to call it transmission 
coefficient
now) may differ from 1 quite a lot.  However, WKB method still
gives a good approximation.  We now assume that WKB condition (3) is
satisfied for $x \leq b$ all the way to $-\infty$. Then,  we only need
to match the outgoing solution $e^{i\sqrt{E}x}$ for $x \geq b$ with
WKB type solutions (2) for $x \leq b$.  Find the coefficient of the
incoming term.  Then the transmission coefficient can be found
\be
T_{\infty} = \frac{4\sqrt{E(E-E_{b})}}{\left[\sqrt{E-E_{b}}+\sqrt{E}
\right]^{2}} \,  . 
\ee
Some values of $T_{\infty}$ are shown in Table~III.  Since both $T_{\infty}$
and $T_{n} \rightarrow 1$ as $E \rightarrow \infty$.  It is more appropriate
to compare reflection coefficients $R_{\infty} \equiv 1-T_{\infty}$
and $R_{n} \equiv 1-T_{n}$.  We see that $R_{\infty}$ gives a very good
approximation to $R_{n}$ for large $E$.  For $E$ close to $E_{b}$, $T_{\infty}$
fails and we need to use $T_{b}$ instead. 

\section*{VIII.  Conclusions}

From the above discussions,  we see that the standard WKB
tunneling factor $T_{w}$ fails badly for $E \rightarrow 0$, $E \approx
E_{b}$ and may have error up to factor of two in between.  However, WKB approximation not necessarily fail provided that we use
appropriate connection relations.  For the example that $k^{2}$ given
by (4),  WKB approximation works for almost entire  energy range
from $E = 0$ to $E \rightarrow \infty$ if we use different
connection relations for different ranges of energy.  Our conclusion
is that when we use the standard WKB formula $T_{w}$,  we need to be
very careful.  If higher accuracy is desired,  we need to consider
connection relations case by case. 

\section*{Appendix:Numerical calculation of $T_{n}$}

Although a numerical method to find $T_{n}$ was described in ref.\ 4, 
a different method was used in this paper. Let
\begin{eqnarray*}
\psi(b) & = & e^{i\sqrt{E}b} \\
\psi_{x}(b) & = & i\sqrt{E}e^{i\sqrt{E}b}, 
\end{eqnarray*}
so that $\psi$ connects to an outgoing solution $e^{i\sqrt{E}x}$
for $x \geq b$. Use Runga--Kutta method to integrate (1), with
$k^{2}$ given by (4),  from $x = b$ back to $x = 0$.  Match with
incoming and reflected waves for $x \leq 0$ and find the coefficient
of the incoming term.  Then the transmission coefficient can be
found by
\begin{eqnarray*}
T_{n} = 4\left/\left|\psi(0)-\frac{i}{\sqrt{E}}\psi_{x}(0)\right. 
\right|^{2}\,  . 
\end{eqnarray*}

\vfill

\section*{Table I.}

\noindent
Comparison of tunneling probabilities, \,  $T_{w}, T_{1}, T_{2}$ and $T_{b}$ 
calculated by WKB method using different approximations with $T_{n}$
calculated numerically, for energy $E$ in the middle range.  
Parameters $\alpha = 0.040965, E_{b}=1.2776$ 
were chosen so that we may compare with Table I of
ref.\ 4.

\vspace{1in}
\begin{tabular}{llllll} \hline\hline
\vspace{-3mm} \\
\baselineskip 25pt
$E$\hspace{11mm} & $T_w$\hspace{16mm} & $T_1$\hspace{16mm} & $T_2$\hspace{16mm} &
$T_b$\hspace{16mm}
 & $T_n$    \vspace{2mm}\\ \hline
\vspace{-3mm} \\
0        & 0.00181  & 0        & 0        & \ \ ---  & 0       \vspace{2mm}\\
0.1      & 0.00611  & 0.00656  & 0.00654  & 0.0342   & 0.00987  \vspace{2mm}\\
0.2      & 0.0144   & 0.0210   & 0.0208   & 0.0543   & 0.0285   \vspace{2mm} \\
0.3      & 0.0297   & 0.0504   & 0.0491   & 0.0896   & 0.0609   \vspace{2mm}  \\
0.4      & 0.0555   & 0.103    & 0.0979   & 0.142    & 0.112    \vspace{2mm} \\
0.5      & 0.0962   & 0.188    & 0.171    & 0.214    & 0.187    \vspace{2mm} \\
0.6      & 0.156    & 0.312    & 0.269    & 0.301    & 0.283    \vspace{2mm} \\
0.7      & 0.240    & 0.479    & 0.382    & 0.397    & 0.391    \vspace{2mm} \\
0.8      & 0.351    & 0.678    & 0.495    & 0.494    & 0.500    \vspace{2mm} \\
0.9      & 0.486    & 0.886    & 0.590    & 0.584    & 0.597    \vspace{2mm} \\
1.0      & 0.640    & 1.06     & 0.647    & 0.663    & 0.678    \vspace{2mm} \\
1.1      & 0.799    & 1.11     & 0.646    & 0.729    & 0.741    \vspace{2mm} \\
1.2      & 0.938    & 0.896    & 0.537    & 0.783    & 0.791    \vspace{2mm} \\
$E_b$    & 1        & 0        & 0        & 0.817    & 0.821
\vspace{2mm} \\ \hline\hline
\end{tabular}

\vfill

\section*{Table II.}

\noindent
Comparison of tunneling probabilities, \,  $T_{w}$, \, $T_{2}$\,  and \,  $T_{0}$, calculated by WKB method using different 
approximations with $T_{n}$ calculated numerically, 
for energy $E \rightarrow 0$\,  ($\alpha = 0.040965$, $E_{b}=1.2776$).

\vspace{1in}
\begin{tabular}{lllll}\hline\hline
\vspace{-3mm} \\
\baselineskip 25pt
$E$ \hspace{10mm} & $T_w$\hspace{18mm}  & $T_{2}$\hspace{18mm}
& $T_0$ \hspace{18mm}  & $T_n$
\vspace{2mm} \\ \hline  \vspace{-3mm} \\
$10^{-8}$&$1.81\ts 10^{-3}$&$6.42\ts 10^{-7}$&$5.97\ts 10^{-10}$&$6.78\ts 10^{-10}$\vspace{2mm}\\
$10^{-7}$&$1.81\ts 10^{-3}$&$2.03\ts 10^{-6}$&$5.97\ts 10^{-9 }$&$6.78\ts 10^{-9}$\vspace{2mm}\\
$10^{-6}$&$1.81\ts 10^{-3}$&$6.42\ts 10^{-6}$&$5.97\ts 10^{-8 }$&$6.78\ts 10^{-8}$\vspace{2mm}\\
$10^{-5}$&$1.81\ts 10^{-3}$&$2.03\ts 10^{-5}$&$5.97\ts 10^{-7 }$&$6.78\ts 10^{-7}$\vspace{2mm}\\
$10^{-4}$&$1.82\ts 10^{-3}$&$6.44\ts 10^{-5}$&$5.99\ts 10^{-6 }$&$6.78\ts 10^{-6}$\vspace{2mm}\\
$10^{-3}$&$1.86\ts 10^{-3}$&$2.08\ts 10^{-4}$&$6.11\ts 10^{-5 }$&$6.80\ts 10^{-5}$\vspace{2mm}\\
$10^{-2}$&$2.17\ts 10^{-3}$&$7.64\ts 10^{-4}$&$7.11\ts 10^{-4 }$&$7.04\ts 10^{-4}$\vspace{2mm}\\
$10^{-1}$&$6.11\ts 10^{-3}$&$6.54\ts 10^{-3}$&$1.93\ts 10^{-2 }$&$9.87\ts 10^{-3}$\vspace{2mm}\\
\hline\hline
\end{tabular}
\vfill

\section*{Table III.}

\noindent
Comparison of tunneling probabilities, \,  $T_{b}$\, and \,  $T_{\iif}$, 
\,  calculated by WKB method using different approximations with $T_{n}$
calculated numerically, for energy $E > E_b$.
$R_{\iif} = 1-T_{\iif} $and $R_{n} = 1 - T_{n} $ are reflection coefficients
($\alpha = 0.040965, E_{b}=1.2776$).

\vspace{1in}
\begin{tabular}{llllll}\hline\hline
\vspace{-3mm} \\
\baselineskip 25pt
$E$ \hspace{10mm} & $T_b$\hspace{15mm}  & $T_{\iif}$\hspace{13mm}
& $T_n$\hspace{15mm} & $R_{\iif}$\hspace{15mm}  & $R_n$
\vspace{2mm} \\ \hline  \vspace{-3mm} \\
1.3      & 0.826    & 0.410    & 0.829   & 0.509     & 0.171 \vspace{2mm}   \\
1.6      & 0.907    & 0.855    & 0.901   & 0.145     & 0.0986\vspace{2mm}   \\
2.0      & 0.956    & 0.938    & 0.947   & 0.0621    & 0.0528  \vspace{2mm} \\
2.6      & 0.985    & 0.972    & 0.974   & 0.0280    & 0.0262  \vspace{2mm} \\
3.0      & 1.01     & 0.981    & 0.982   & 0.0190    & 0.0184  \vspace{2mm} \\
4.0      & \ \ ---  & 0.991    & 0.991   & 0.00920   & 0.00901 \vspace{2mm} \\
5.0      & \ \ ---  & 0.995    & 0.995   & 0.00542   & 0.00539 \vspace{2mm} \\
6.0      & \ \ ---  & 0.996    & 0.996   & 0.00357   & 0.00356 \vspace{2mm} \\
7.0      & \ \ ---  & 0.997    & 0.997   & 0.00253   & 0.00252 \vspace{2mm} \\
\hline\hline
\end{tabular}
\vfill
\section*{References}
\newref{1}
H.\ Jeffreys, ``On  certain approximate solutions of linear
differential equations of the second order, '' Proc.\ London Math.\
Soc.\ {\bf 23} 428--436\ (1923).
\newref{2}
G.\ Wentzel, ``A generalization of quantum conditions for the 
purposes of wave mechanics,''
Zeits.\ f.\ Physik, \ {\bf 38}\ 518--529\ (1926).
H.\ A.\ Kramers, ``Wave mechanics and semi-numerical quantisation,''
Zeits.\ f.\ Physik, \ {\bf 39}\ 828--840\ (1926).
L.\ Brillouin, ``The undulatory mechanics of Schr\"{o}dinger,''
Comptes Rendus, {\bf 183} 24--26\ 270--271\ (1926).
L.\ Brillouin, ``Notes on undulatory mechanics,''
J.\ de Physique et le Radium, {\bf 7}\ 353--368\ (1926).
\newref{3}
D.\ G.\ Swanson, {\em Plasma Waves \/}(Academic, San Diego, CA, 1989), pp.\ 13--15, pp.\ 234--238.
\newref{4}
J.\ Crofton,  P.\ A.\ Barnes, and M.\ J.\ Bozack, ``Quantum mechanical tunneling 
in an ohmic contact,''Am.\ J.\ Phys., {\bf 60}\ 499--502\ (1992).
\newref{5}
H.\ Enge, \ {\em Introduction to Nuclear Physics\/}\ (Addison--Wesley, 
Reading, \,  MA, \,  1966),  pp.\ 282--291.
\newref{6}
L.\ D.\ Landau and E.\ M.\ Lifshitz, {\em Quantum Mechanics \/}(Pergamon, 
Oxford, \,  England, \,  1977), 3rd ed.,  \  pp.\ 178--181.
\newref{7}
R.\ H.\ Dicke and J.\ P.\ Wittke, {\em Introduction to Quantum Mechanics\/}  
(Addison--Wesley, Reading, MA, 1960), pp.\ 245-253.
\newref{8}
E.\ Merzbacher, {\em Quantum Mechanics\/}(Wiley, New York, 1970), 2nd ed., 
pp.\ 116--138.
\pagebreak
\newref{9}
E.\ Segr\`{e}, {\em Nuclei and Particles\/}\,  
(W.\ A.\ Benjamin, Reading, MA, 1977),  2nd ed.,  pp.\ 319--328.
\newreff{10}
M.\ Abramowitz and I.\ A.\ Stegun, \ {\em Handbook of Mathematical 
Functions \/}(National Bureau of Standards, Washington D.\ C.,  1964).
\newreff{11}
With the knowledge of parabolic cylinder functions,  it can be shown
that $\mbox{Aj$'(0)$} = -2\pi^{1/2}/\Gamma (1/4) \approx -0.97774$.
\vfill
\end{document}